\documentclass[aps,prb,reprint,superscriptaddress,longbibliography]{revtex4-1}
\usepackage{times}
\usepackage{graphicx}
\usepackage{hyperref}
\usepackage{amsmath}

\usepackage{xcolor}
\newcommand{\Revision}{\textcolor{black}}

\begin{document}


\title{Non-local transport measurements in hybrid quantum Hall - superconducting devices}

\author{Lingfei Zhao}
\email{lz117@duke.edu}
\affiliation{Department of Physics, Duke University, Durham, NC 27708, USA}

\author{Ethan G. Arnault}
\affiliation{Department of Physics, Duke University, Durham, NC 27708, USA}

\author{Trevyn F.Q. Larson}
\affiliation{Department of Physics, Duke University, Durham, NC 27708, USA}


\author{Kenji Watanabe}
\affiliation{National Institute for Materials Science, 1-1 Namiki, Tsukuba 305-0044, Japan}

\author{Takashi Taniguchi}
\affiliation{National Institute for Materials Science, 1-1 Namiki, Tsukuba 305-0044, Japan}

\author{Fran\c cois Amet}
\affiliation{Department of Physics and Astronomy, Appalachian State University, Boone, NC 28607, USA}

\author{Gleb Finkelstein}
\affiliation{Department of Physics, Duke University, Durham, NC 27708, USA}

\date{\today}


\begin{abstract} 
There has been a growing interest in hybrid quantum Hall (QH) – superconductor devices, driven by the prospect to realize exotic ground states and excitations with non-abelian exchange statistics. While the existing experiments clearly demonstrate Andreev coupling between the edge states and the superconductors, the question remains whether the quantum coherence could propagate between several superconducting contacts via chiral channels. To answer this question, we have first extended the Landauer-B\"uttiker (LB) formalism to samples with one superconducting contact and found a remarkable agreement within a series of measurements related to each other via LB-type formulae. We have then switched to the case of multiple superconducting contacts, and found that we can describe the measurements self-consistently if we neglect the superconducting phase coherence between multiple contacts. We interpret this result as a negative answer to the question posed above: the phase correlations between multiple superconducting contacts are not established via micron-long quantum Hall edge states. Looking forward, our approach may find applications in the broader field of topological superconductivity and proximal structures. Possible violations of the self-\Revision{consistency} tests presented here may be used as an indication that superconducting phase coherence is induced in the quantum Hall edges.
\end{abstract}
\maketitle  


There has been a long-standing interest to inducing superconducting correlations into the quantum Hall edge states~\cite{Ma1993,Fischer1994,takagaki_transport_1998,Takayanagi1998,Moore1999,Hoppe2000,Eroms2005,Giazotto2005,Batov2007,Akhmerov2007,Chtchelkatchev2007,Rakyta2007,Khaymovich2010,Stone2011,vanOstaay2011,Komatsu2012,Rickhaus2012,SanJose2015,Wan2015,Liu2017,Sahu2018,Beconcini2018,Alavirad2018,Zhangsongbo2019,PeraltaGavensky2020}. The recent experimental developments were enabled by the non-local transport measurements, which allow one to explore the properties of the hybrid quantum Hall (QH) -- superconductor interfaces~\cite{Lee2017,Park2017,Kozuka2018,Matsuo2018,Zhao2020,DaWang2021,Gl2022,Hatefipour2022,Hatefipour2023}, thus attracting significant theoretical attention~\cite{Manesco_mechanisms_2021,Kurilovich_disorder_2022,Michelsen_supercurrent_2022,Tang_2022,David_effects_2022,Schiller_interplay_2022}. While QH samples with normal contacts are conventionally described by the Landauer-B\"uttiker (LB) formalism~\cite{Beenakker1997}, the situation could be more complicated in the case of hybrid superconducting devices, as both electrons and reflected holes may have to be considered coherently~\cite{Lambert1998}. 
\Revision{In principle, one can account for the superconducting correlations between multiple superconductors using the scattering matrix in the particle-hole Nambu space obtained from Bogoliubov-de Gennes (BdG) equations~\cite{Lambert1993}. However, the calculations of the nonlocal resistance in the QH-superconductor structures are commonly simplified by using the LB expressions and describing the superconducting contact via two parameters: the probabilities of the normal and Andreev reflections~\cite{takagaki_transport_1998}.}

\Revision{Here, we follow the latter approach and derive a simple extension of the LB formalism suitable for the QH samples with multiple superconducting contacts, applicable in the case when coherence between the contacts can be neglected.} Namely, we assume that the effect of the superconducting contacts is limited to setting the total electron occupation of the QH channels flowing downstream, and we do not track the Andreev-reflected holes separately from the normally reflected electrons. This approach is based on the expectation that only the total current flowing downstream from the superconductor is relevant for the non-local small-bias differential resistance, as we will discuss in the following. We therefore characterize each superconducting contact by its ``generalized reflection coefficient'', which ranges from 1 to $-1$, where these limits corresponds to a perfect normal or Andreev reflection of the electrons arriving at the contact.
We do not attempt to evaluate this coefficient, which is determined by the microscopic details of the proximitized graphene interface. Instead, we extract it from the experiments for consistency checks.  
 
We test our results on actual devices with multiple superconducting contacts or QH channels. By measuring several non-local resistance configurations, we can verify the internal consistency of our expressions. The good agreement with the experiment indicates that the neglected coherence effects are either small or should not exist in the present samples. Finally, we discuss possible breakdown of our approach when superconducting phase coherence between contacts becomes important.

\section*{A superconductor contacting quantum Hall edge states}

In the absence of coherence between the superconducting contacts, we can treat them as incoherent reservoirs and use the standard LB formalism to calculate the current flowing into a lead connected to the superconductor. We start by considering a multi-terminal device with only one conducting channel in each lead. (Later we generalize the results to the case of multiple channels.) The standard expression for the current going into lead $\alpha$ ($\alpha=$ 1, 2, ...) is
\begin{equation}
I_\alpha=\frac{e}{h}\sum_\beta\int  A_{\beta \beta}^\alpha(E)  n_\beta (E) dE.
\label{currenteq2}
\end{equation}
Here, $A_{\beta \gamma}^\alpha=\delta_{\alpha \beta}\delta_{\alpha \gamma}-s^\dagger_{\alpha \beta} s_{\alpha \gamma}$, where $s_{\alpha\beta}$ is the scattering matrix from lead $\beta$ to $\alpha$, which describes the QH part of the sample excluding the contacts. $n_\beta (E)$ is the distribution function of the channel leaving contact $\beta$ at energy $E$. Note: we use ``arriving'' and ``leaving'' when describing the direction of particles with respect to the \emph{contacts}. This way we intend to avoid the confusion with ``outgoing'' and ``incoming'', which conventionally refer to the direction of electrons with respect to the interior of the device. 

In equilibrium, the channel occupation number $n_\beta(E)$ at zero-temperature is a step function with electrons filled to $\mu_\beta$, the chemical potential of the channel leaving lead $\beta$, and for a small excitation near the Fermi level, Eq.~\ref{currenteq2} reduces to 
\begin{equation}
\delta I_\alpha=\frac{e}{h}\sum_\beta A_{\beta\beta}^\alpha (E_F) \delta \mu_\beta.
\label{dIvsdE}
\end{equation}
For normal metal contacts, the chemical potential of electrons leaving them is in equilibrium, and we have $\mu_\beta=e V_\beta$, where $V_\beta$ is the bias voltage applied to contact $\beta$. 

For superconducting contacts, the situation is more complicated due to the Andreev reflections: the occupation of the channel leaving contact $n_\beta$ is not solely determined by $V_\beta$ but has contributions from the upstream contacts. As a result, the distribution  downstream of the superconducting contact may be different from equilibrium. However, in our simplified approach, we describe this occupation by an equilibrium distribution function which contains the same number of electrons. 

This crucial step is at the core of our approach. It clearly does not change the results if the downstream contact is made of a normal metal, which absorbs and equilibrates all the arriving electrons, so that the current flowing into that contact is correctly accounted for. Later on, we argue that this procedure can also be used with some limitations in the case where several superconducting contacts are present. 
Finally, in the multichannel case, each channel emerging from a superconducting contact will be characterized by its own effective chemical potential, depending on the contributions of the normal and Andreev reflections to that particular channel.

\begin{figure}
\centering
\includegraphics[width=0.7\columnwidth]{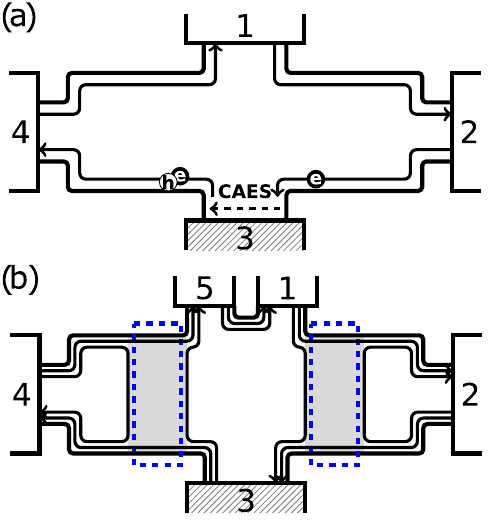}
\caption{Sketches of QH devices with one superconducting contact and multiple normal contacts. (a) The device is in QH regime with only one QH edge state propagating clockwise. The shaded contact (3) is superconducting and others are normal. CAES along the superconducting contact convert an upstream electron into a downstream electron or hole. (b) Two QH edge states are present (e.g. belonging to different Landau levels.) The inner channels are reflected back by the gates (blue dotted rectangles).}
\label{Fig1}
\end{figure}

As an example, we first demonstrate the implementation of Eq.~\ref{dIvsdE} in a Hall bar geometry shown in Fig.~\ref{Fig1}(a). The device has three normal contacts (1, 2 \& 4) and one superconducting contact (3). As discussed before, the chemical potential of the edges states downstream of the normal contacts ($\beta=1,2,4$) is equal to the applied bias: 
\begin{equation}
 \mu_{\beta=1,2,4}=e V_\beta.
\label{NE}
\end{equation} 
For superconducting contact 3, an electron arriving from the upstream contact 2 can leave as a normally reflected electron with a probability of $P_e$, or as an Andreev-reflected hole with a probability of $P_h$. The electron can also be absorbed and emitted by the subgap states of the non-ideal superconductor, so $P_e +P_h$ does not have to be equal to 1. These probabilities are determined by the microscopic properties of the superconductor and their properties are being actively explored. In the following, these probabilities will be extracted from the experiment.

The distribution flowing downstream from contact 3 has three contributions: the normally reflected electrons and Andreev reflected holes originating from contact 2, and the electrons originating from the subgap states of contact 3: 
\begin{equation}
\begin{split}
n_3(E)&=P_e f_0(E-\mu_2)+P_h f_0(E - (2eV_3 - \mu_2)) \\ 
&+ (1-P_e-P_h)f_0(E-eV_3),
\label{eff_distr}
\end{split}
\end{equation}
where $f_0(E-\mu)$ is the Fermi occupation with a chemical potential $\mu$. 

We approximate $n_3(E)$ by an effective distribution $f_0(E-\mu_3)$, where $\mu_3$ is chosen so that the two distribution would contain the same number of electrons: $\int f_0(E-\mu_3)dE = \int n_3(E)dE$. This yields $\mu_3=(1-P_e+P_h) eV_3+(P_e-P_h)\mu_2$. As argued above, $f_0(E-\mu_3)$ correctly describes the current flowing from contact 3 to contact 4.

We next introduce the \emph{generalized reflection coefficient}, $P=P_e-P_h$, defined as the difference between the normal and Andreev reflection probabilities. The expression for the chemical potential then simplifies to 
\begin{equation}
\mu_3=(1-P) eV_3+P\mu_2,
\label{1SCmuV}
\end{equation} 
where the first term describes the electrons originating from contact 3, while the second term corresponds to the normal and Andreev reflections of the electrons arriving from contact 2. We further rewrite this expression as 
\begin{equation}
\delta \mu_3=e \delta V_3 + eP(\delta V_2-\delta V_3).
\label{SCE}
\end{equation}

The device is in the quantum Hall (QH) regime with an edge state propagating clockwise. The scattering matrix of the QH region gives 
\begin{equation}
A_{\beta\beta}^\alpha=\delta_{\alpha,\beta}-\delta_{\alpha-1, \beta}.
\label{A1SC}
\end{equation} 
Here, we use circular indexing, i.e. index 0 refers to contact 4 and index 5 refers to contact 1. 

Plugging Eq.~\ref{A1SC},~\ref{NE}~\&~\ref{SCE} into Eq.~\ref{dIvsdE} gives 
\begin{gather}
\begin{pmatrix}
\delta I_1 \\ \delta I_2 \\ \delta I_3 \\ \delta I_4
\end{pmatrix}=\frac{e^2}{h}
\begin{pmatrix}
 1 & 0 & 0 & -1 \\ -1 & 1 & 0 & 0 \\ 0 & -(1-P) & 1-P & 0 \\ 0 & -P & -(1-P) & 1  
\end{pmatrix}
\begin{pmatrix}
\delta V_1 \\ \delta V_2 \\ \delta V_3 \\ \delta V_4
\end{pmatrix}.
\label{matrix1}
\end{gather}
The sum of the elements of the conductance matrix in each column and row is zero, confirming current conservation and the absence of current flow in equilibrium. This result has been previously presented in the supplementary material to our experiment~\cite{Zhao2020}.

In experiments, one typically inject current excitation, $\delta I$, from one contact to another while measuring the voltage responses from which the differential resistances can be calculated. We define the differential resistance between a pair of contacts $i$ and $j$ while flowing $\delta I$ from contact $k$ to $l$ as 
\begin{equation}
R_{lk,ij}=\frac{\delta V_i - \delta V_j}{\delta I}.
\end{equation}  

As an example, we consider injecting current at contact 1 while grounding the superconducting contact. The boundary condition is $(\delta I_1, \delta I_2, \delta I_3, \delta I_4)=(\delta I, 0, -\delta I,0)$. Solving Eq.~\ref{matrix1}, we obtain the Hall resistance, $R_H$, and the downstream resistance, $R_D$:
\begin{align}
R_H&=R_{31,24}=\frac{h}{e^2},\\
R_D&=R_{31,43}=R_H\frac{P}{1-P}.
\label{RDvsP}
\end{align}
$R_H$ does not depend on $P$ and remains quantized. $R_D$ deviates from zero and has a nonlinear dependence on $P$. The sign of $R_D$ is the same as $P$, i.e. positive if the  particles flowing downstream are mostly electrons and negative if holes dominate. The maximum of $R_D$ is infinity, corresponding to an insulating interface. The minimum $R_D$ is equal to $-R_H/2$, agreeing with perfect Andreev conversion,  which doubles the interface conductance. Oscillations of $R_D$ around zero together with quantized $R_H$  while tuning $P$ have been observed in previous experiments~\cite{Zhao2020,Zhao2022LossDecoh}. Note that Eq.~\ref{RDvsP} provides a convenient way of determining $P$ in experiments, i.e. 
\begin{equation}
P=\frac{R_D}{R_D+R_H}.
\end{equation}

\section*{Multiple edge states}

When there are multiple Landau levels, we need to consider the current contribution from each corresponding edge state $i$ separately. Eq.~\ref{dIvsdE} becomes
\begin{align}
\delta I_\alpha =& \sum_i \delta I_{\alpha i}\\
\delta I_{\alpha i}=&\frac{e}{h}\sum_{\beta j} g_j A_{\beta j \beta j}^{\alpha i} \delta \mu_{\beta j},
\label{dIvsdE2}
\end{align}
where the tensor $A_{\beta j \beta j}^{\alpha i} $ is obtained via the scattering matrices,
\begin{equation}
A_{\beta j \gamma k}^{\alpha i}=\delta_{\alpha \beta}\delta_{\alpha \gamma} \delta_{i j} \delta_{i k} - s_{\alpha i, \beta j}^{\dagger} s_{\alpha i, \gamma k}
\end{equation}
Here, indices $\alpha$ and $\beta$ refers to the leads, $i$, $j$ refer to the Landau levels, and $g_j$ is the degeneracy of Landau level $j$. Note that the chemical potentials $\mu_{\beta j}$ of the different edge states $j$ leaving the same contact $\beta$ are assumed to be generally different from each other.

To illustrate the effect of a superconductor contacting multiple Landau levels, we consider a device shown in Fig~\ref{Fig1}(b). Contact 3 is a superconductor and others are normal metals. For simplicity, we consider the lowest two Landau levels of graphene, 
where outer channel corresponds to the first Landau level with the degeneracy $g_1=2$ and the inner one corresponds to the second Landau level with $g_2=4$. The electron density in the gray regions is lower, allowing only the outer edge to transmit. In this geometry, Eq.~\ref{dIvsdE2} gives

\begin{widetext}
\begin{gather}
\resizebox{0.06\columnwidth}{!}{$
\begin{pmatrix}
\delta I_1 \\ \delta I_2 \\ \delta I_3 \\ \delta I_4 \\ \delta I_5
\end{pmatrix}$}=\frac{e}{h}
\resizebox{0.7\columnwidth}{!}{$
\begin{pmatrix}
g_1 \delta \mu_{11} + g_2 \delta \mu_{12} & 0 & 0 & 0 & -(g_1 \delta \mu_{51}+g_2 \delta \mu_{52})\\
-g_1 \delta \mu_{11} & g_1 \delta \mu_{21} & 0 & 0 & 0\\
-g_2 \delta \mu_{12} & -g_1 \delta \mu_{21} & g_1 \delta \mu_{31} + g_2 \delta \mu_{32} & 0 & 0\\
0 & 0 & -g_1 \delta \mu_{31} & g_1 \delta \mu_{41} & 0\\
0 & 0 & -g_2 \delta \mu_{32} & -g_1 \delta \mu_{41} & g_1\delta \mu_{51}+g_2\delta \mu_{52}
\end{pmatrix}
$},
\label{matrix2}
\end{gather}
\end{widetext}

\begin{figure*}[t!]
\centering
\includegraphics[width=2\columnwidth]{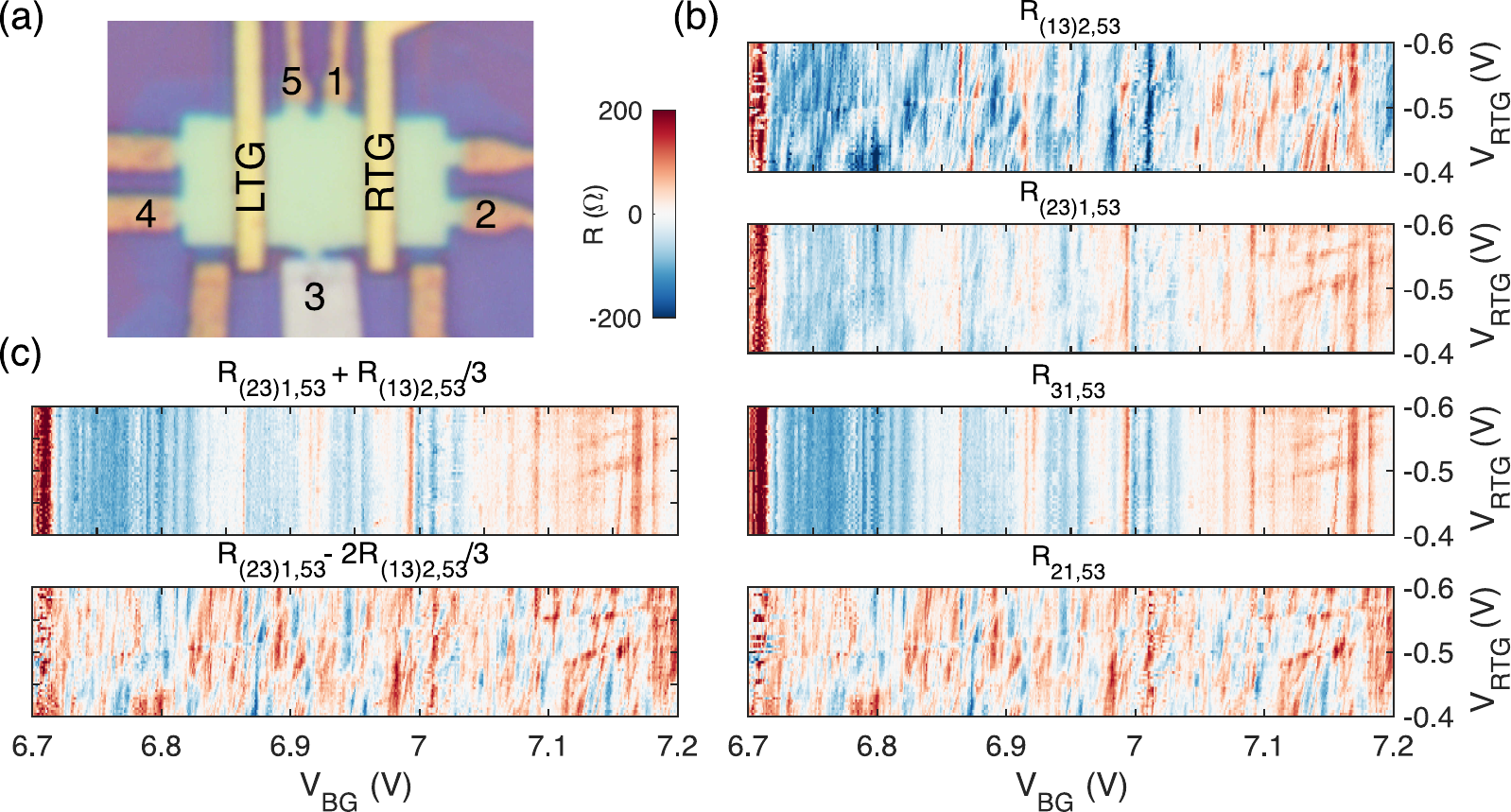}
\caption{(a) Optical image of the device. The filling factor under the top gates (LTG and RTG) is $\nu=2$. The filling factor outside is tuned to $\nu=6$ by a global Si gate (BG) underneath the substrate. Contact 3 is a superconductor (MoRe) and other contacts are normal metals (Cr/Au). (b) $R_{(13)2,53}$, $R_{(23)1,53}$, $R_{31,53}$ and $R_{21,53}$ (from top to bottom) measured as a function of $V_{RTG}$ and $V_{BG}$. $V_{BG}$ is tuned inside the range of $\nu=6$ plateau and $V_{RTG}$ is maintaining $\nu=2$ underneath. (c) $R_{(23)1,53}+R_{(13)2,53}/3$ (top) and $R_{(23)1,53}-2R_{(13)2,53}/3$ (bottom) plotted vs. $V_{RTG}$ and $V_{BG}$. The two maps agree well with $R_{31,53}$ and $R_{21,53}$ in (b).}
\label{Fig2}
\end{figure*}

where $\delta \mu_{\alpha \neq 3,i}=e \delta V_\alpha$ for the normal contacts. For the superconducting contact, the generalized reflection coefficient is a $2\times 2$ matrix, $\textbf{P}=(P_{ij})$, where $P_{ij}$ is the difference between the normal and Andreev reflection probabilities from Landau level $j$ to $i$. For convenience, we define the total generalized reflection coefficient for an electron initially on Landau level j as $P_j=\sum_i P_{ij}$. Considering various normal and Andreev reflection processes, Eq.~\ref{SCE} becomes
\begin{equation}
\delta \mu_{3i} = e\delta V_3 + P_{i1} (\delta \mu_{21} - e \delta V_3) + P_{i2}(\delta \mu_{12} -e \delta V_3).
\end{equation}
Plugging the expression of $\delta \mu_{\alpha i}$ into Eq.~\ref{matrix2} gives
\begin{widetext}
\begin{gather}
\resizebox{0.06\columnwidth}{!}{$
\begin{pmatrix}
\delta I_1 \\ \delta I_2 \\ \delta I_3 \\ \delta I_4 \\ \delta I_5
\end{pmatrix}$}=\frac{e^2}{h}
\resizebox{0.65\columnwidth}{!}{$
\begin{pmatrix}
g_1   + g_2   & 0 & 0 & 0 & -(g_1 +g_2)\\
-g_1  & g_1  & 0 & 0 & 0\\
-g_2(1-P_{2})  & -g_1(1-P_{1})  & g_1+g_2-g_1P_1-g_2P_2   & 0 & 0\\
-g_2P_{12} & -g_1P_{11} & -g_1(1-P_{11})+g_2P_{12}  & g_1  & 0\\
-g_2P_{22} & -g_1P_{21} & -g_2(1-P_{22})+g_1P_{21}  & -g_1  & g_1+g_2
\end{pmatrix}$}
\resizebox{0.06\columnwidth}{!}{$
\begin{pmatrix}
\delta V_1 \\ \delta V_2 \\ \delta V_3 \\ \delta V_4 \\ \delta V_5
\end{pmatrix}$}.
\label{matrix3}
\end{gather}
\end{widetext}

We first look at the voltage response at contact (5) while leaving contact (4) floating. In this case, all the particles coming out of the superconducting contact (3) are collected by (5). We are interested in measuring the downstream voltage $V_{53}$ in the following biasing configurations: 
\begin{enumerate}
\item \Revision{$R_{31,53}$}: injecting current from  (1) to grounded (3); 
\item \Revision{$R_{21,53}$: injecting current from  (1) to grounded (2);}
\item \Revision{$R_{(13)2,53}$}: injecting from (2) while grounding both (1) and (3); 
\item \Revision{$R_{(23)1,53}$}:  injecting from (1) while grounding both (2) and (3). 
\end{enumerate}
Note that in order to indicate multiple grounded contacts, we group them in parenthesis in the subscript. 
Solving Eq.~\ref{matrix3} with corresponding boundary conditions, we obtain
\begin{align}
R_{(13)2,53} &= \frac{h}{6e^2} P_1 \label{R531}\\
R_{(23)1,53} &= \frac{h}{6e^2} \frac{2P_2}{3-2P_2} \\
R_{31,53} &= \frac{h}{6e^2} \frac{P_1+2P_2}{3-(P_1+2P_2)}\\
R_{21,53} &= \frac{h}{6e^2} \frac{2(P_2-P_1)}{3-(P_1+2P_2)}\label{R534}.
\end{align}

To examine these relations, we fabricate a graphene device (see Fig~\ref{Fig2}(a)) to realize the configuration in Fig.~\ref{Fig1}(b). The superconducting contact 3 is made of sputtered MoRe alloy and the normal contacts are made of thermally evaporated Cr/Au. The device is in the QH regime at an external magnetic field of 2.5 T. We apply a positive voltage, $V_{BG}$, to the Si gate (BG) underneath the SiO$_2$ insulator to tune the bulk of the device to a filling factor $\nu=6$. Negative voltages $V_{RTG}$ and $V_{LTG}$ are then applied to the local top gates (LTG and RTG) to deplete the regions underneath to a filling factor $\nu=2$. The temperature of the device is 40 mK. 

$R_{(13)2,53}$, $R_{(23)1,53}$, $R_{31,53}$ and $R_{21,53}$ measured in this device are plotted in Fig.~\ref{Fig2}b as a function of $V_{RTG}$ and $V_{BG}$. 
The detailed understanding of the observed signal dependencies on the gate voltages is not required. However, qualitatively, BG controls the $P_{eh}$ patterns of the first and second Landau levels, while RTG controls the mixing of these two Landau levels before they arrive at the SC contact. All the configurations chosen here do not depend on the LTG. In the following, we cross-check the patterns presented in  Fig.~\ref{Fig2}b in order to check the validity of our analysis.

For all the four differential resistances, $V_{BG}$ strongly affects the oscillation patterns. This is because $V_{BG}$ directly tunes the electron density nearby the superconductor and therefore affects $P_{eh}$. In contrast, $V_{RTG}$ locally controls the electron density underneath the top gate and has no effect on the electron density near the superconductor. Naively, one would expect that $V_{RTG}$ should not influence the oscillation patterns. However, this is only true for $R_{31,53}$ (third panel in Fig.~\ref{Fig2}b), where the inner and outer edge states are in equilibrium before reaching the superconductor. In the other biasing schemes, the two edge states have different chemical potentials and their scattering, influenced by $V_{RTG}$, affects the results of the measurement. 

Inter-Landau level scattering mixes the various scattering channels together and hinders the characterization of individual $P_{ij}$. However, Eq.~\ref{R531}-\ref{R534} are still effective if we include the effects of inter-Landau level scattering into the definition of the generalized reflection coefficients, effectively making $P_{ij}$ dependent on $V_{RTG}$.

Note that $R_{(13)2,53}$, $R_{(23)1,53}$, $R_{31,53}$ and $R_{21,53}$ are all much smaller than $h/6e^2$, suggesting $|P_{1,2}|\ll1$, likely due to the electron absorption by the superconducting vortices~\cite{Zhao2022LossDecoh}. In this case, Eq.~\ref{R531}-\ref{R534} approximates to $R_{(13)2,53}\approx(h/6e^2) P_1$, $R_{(23)1,53}\approx(h/6e^2)2P_2$, $R_{31,53}\approx(h/6e^2)(P_1+2P_2)/3$ and $R_{21,53}\approx(h/6e^2)2(P_2-P_1)/3$. Therefore, we have 
\begin{align}
R_{31,53}\approx&\ R_{(23)1,53}+\frac{R_{(13)2,53}}{3}\\
R_{21,53}\approx&\ R_{(23)1,53}-\frac{2R_{(13)2,53}}{3}.
\end{align}
The quantities at the right side of the above equations are directly extracted from the two top panels of Fig.~\ref{Fig2}b and are plotted in Fig.~\ref{Fig2}c. They should be compared to the two bottom panels of Fig.~\ref{Fig2}c representing $R_{31,53}$ and $R_{21,53}$ obtained in the experiment. Indeed the two pairs of maps (Fig.~\ref{Fig2}c and the bottom two in Fig.~\ref{Fig2}b) show an excellent agreement in the full range of gate voltages.

\begin{figure*}[ht!]
\centering
\includegraphics[width=1.5\columnwidth]{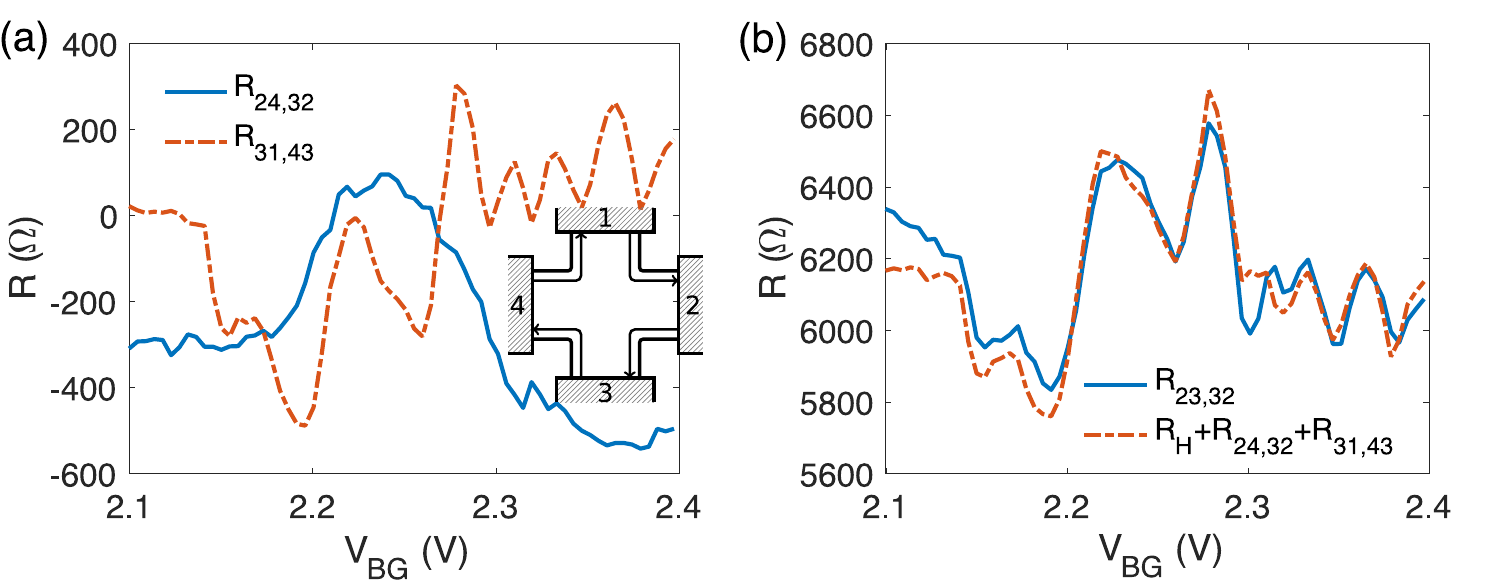}
\caption{(a) $R_{24,32}$ and $R_{31,43}$ of a bilayer graphene device with 4 superconducting contacts (inset) plotted vs. $V_{BG}$. The gate voltage spans over the lowest Landau level ($\nu=4$) plateau. (b) $R_{23,32}$ vs. $R_H+R_{24,32}+R_{31,43}$ as a function of $V_{BG}$. The two differential resistances agree well. \Revision{The temperature of the device is 60 mK.}}
\label{Fig3}
\end{figure*}

\section*{Multiple superconducting contacts}

Next, we consider the case of several superconducting contacts. For simplicity, we consider only one QH edge state. We describe all the scattering processes on the level of probabilities, thereby neglecting any coherent effects involving multiple contacts. However, within this approximation, the successive reflections from multiple contacts are treated accurately. An example of a step-by-step consideration for the case of two successive superconducting contacts is presented in the appendix.

Here, we start with Eq.~\ref{1SCmuV} which is generalized to multiple superconducting contacts as
\begin{equation}
\mu_{\alpha}=(1-P_{\alpha})eV_{\alpha}+P_{\alpha}\mu_{\alpha-1}.
\label{AllSCdVdE}
\end{equation}
This expression allows us to directly relate $\mu_{\alpha-1}$ and $\mu_{\alpha}$ -- the effective chemical potentials upstream and downstream of contact $\alpha$ -- without having to track the chains of reflections from multiple contacts.

We next consider a device in which all contacts are  superconducting. Eq.~\ref{A1SC} is still valid, i.e. 
\begin{equation}
\delta I_\alpha = \frac{e}{h} (\delta \mu_\alpha - \delta \mu_{\alpha-1}).
\end{equation}
Eliminating $\delta \mu_{\alpha-1}$ between the last two equations, we have
\begin{equation}
e\delta V_\alpha = \delta \mu_\alpha + \frac{P_\alpha}{1-P_\alpha} \frac{h}{e} \delta I_\alpha.
\label{AllSCdV}
\end{equation}
For a floating contact $\alpha$, $\delta I_\alpha=0$ results in $e\delta V_\alpha = \delta \mu_\alpha$. Note that this is the property of a normal contact! A superconducting contact can thus be used as a good voltage probe (infinite impedance) in the QH regime if it is contacting a single edge state (or multiple mutually equilibrated edge states). This means, once again, that the Hall resistance remains quantized, $R_H=h/e^2$, even if it's measured between a pair of superconducting contacts. Eq.~\ref{AllSCdV} also suggests that a superconducting contact $\alpha$ of $P_\alpha=0$ effectively behaves like a normal contact. All the results obtained in the following are then applicable to hybrid devices by simply setting $P_\alpha=0$ if $\alpha$ is normal.

We first calculate the two-terminal differential resistances, $R_{ji,ij}$ ($i\neq j$). The boundary condition is $\delta I_i=\delta I$, $\delta I_j=-\delta I$ and $\delta I_{\alpha}=0$ for $\alpha \neq i,j$. Since the floating contacts have $\delta \mu_\alpha=\delta \mu_{\alpha-1} (\alpha \neq i,j)$, we obtain from Eq.~\ref{AllSCdVdE} for biased contacts $i,j$:
\begin{align}
e\delta V_i &= \frac{\delta \mu_i - P_i \delta \mu_{j} }{1-P_i} \\
e\delta V_j &= \frac{\delta \mu_j - P_j \delta \mu_{i} }{1-P_j}.
\end{align} 
In combination with Eq.~\ref{AllSCdV}, we solve 
\begin{equation}
R_{ji,ij}=R_H\frac{1-P_i P_j}{(1-P_i)(1-P_j)}.
\label{R2p}
\end{equation}
Interestingly, the two-terminal resistance continuously approaches zero as the Andreev conversion of both contacts nears unity, i.e. $P_i, P_j \rightarrow -1$. We emphasize that this is not a Josephson effect~\cite{HuangNazarov2017,HuangNazarov2019}, as the phases of the contacts are assumed to be uncorrelated.
\Revision{Instead, the vanishing resistance indicates that in this limit the dissipation is turned off. The absence of dissipation here has two microscopic reasons. First, the absence of backscattering in the QH regime makes sure that there is no equilibration between the carriers traveling in opposite directions along the two different QH edges. Second, the perfect Andreev conversion at the QH-S interfaces ensures that there is no dissipation at these intersections.}

Assuming $i\neq j+1$, the voltage of the contact at the downstream of the grounded contact $j$ is, $e\delta V_{j+1}=\delta \mu_{j+1}=\delta \mu_j$. Eq.~\ref{AllSCdV} gives $e(\delta V_{j+1}-\delta V_j)=\delta I (h/e)P_j/(1-P_j)$, from which we obtain the downstream resistance
\begin{equation}
R_{ji,j+1\ j}=R_H\frac{P_j}{1-P_j},
\label{RDS}
\end{equation}
recovering the results of Eq.~\ref{RDvsP}. Note that the downstream resistances do not depend on which contact is injecting current, providing a convenient way of characterizing the generalized reflection coefficient $P$ of each superconducting contact separately.

To check the results of Eq.~\ref{R2p} \& \ref{RDS}, we studied a sample with 4 superconducting contacts made of a bilayer graphene (see inset of Fig.~\ref{Fig3}(a)). We apply an external magnetic field of 2 T and study the QH state with $\nu=4$, where $R_H=h/4e^2$. The electron density of the device is controlled by the voltage, $V_{BG}$, applied to the Si back gate (BG). Fig.~\ref{Fig3} (a) plots two downstream resistances $R_{24,32}$ and $R_{31,43}$ measured as a function of $V_{BG}$. 
The observed downstream resistances are much smaller than $R_H$, suggesting $|P_2|, |P_3|\ll1$. We then have 
\begin{align}
R_{24,32}&\approx R_H P_2, R_{31,43}\approx R_H P_3 \\
R_{23,32}&\approx R_H (1+P_2+P_3) \approx R_H + R_{24,32} + R_{31,43}. 
\end{align}

In Fig.~\ref{Fig3}(b), we plot $R_{23,32}$ together with $R_H + R_{24,32} + R_{31,43}$ and indeed, they agree well with each other over the full range of $V_{BG}$.  



\section*{Outlook}
The results presented in this paper show that our approach captures the main features of the representative non-local resistance measurements in hybrid superconductor-QH samples. In the following, we discuss future experiments that could either reveal the limitation of our approach or further probe the limits of its validity.  

At the core, our approach neglects the phases of superconductors for a certain type of non-local resistance measurements. This assumption allows us to treat the superconducting electrodes in a similar way to normal electrodes and to combine their generalized reflection probability with the scattering matrix of the normal QH region. In contrast to generic hybrid superconducting structures~\cite{Beenakker1997}, this assumption is rather natural for QH systems with multiple electrodes. Indeed, in the absence of backscattering, the information about the phase of the superconductor is typically lost in the downstream contacts. This is especially true when the downstream contacts are normal (Fig.~1). However, even when only superconducting contacts are present (Fig.~3), the phase information would have to travel around the circumference of the sample to return back to the origin. 
Specifically, the phases of the particles flowing in and out of the superconducting contacts are assumed to be not correlated, in the style of the BTK approximation~\cite{Blonder1982}.  

There are still several scenarios in which the superconducting phase coherence could matter due to the interference effects. Most straightforwardly, such interference can occur if the superconducting contact is placed along the edge of a QH interferometer, so that following the Andreev or normal reflection the particles would loop back to the same contact~\cite{Khrapai2023}. We are currently exploring the coherence length of the chiral Andreev edge states (CAES) propagating along the QH edge states~\cite{Zhao2022LossDecoh}. Placing a superconducting contact inside a QH interferometer would be a natural step for testing the phase coherence of the CAES. The effect of superconducting vortices on the phase coherence of these states is particularly intriguing~\cite{Tang_2022,Kurilovich_disorder_2022,Schiller_interplay_2022}.

Finally, for a QH structure with two superconducting electrodes, the appearance of supercurrent clearly violates the assumptions of this paper~\cite{Amet2016,Seredinski2019,Vignaud2023}. In samples with more than two contacts, the supercurrent could be induced by propagating the phase information either outside of chiral channels~\cite{HuangNazarov2017}, or around the circumference of the sample as mentioned earlier.

\begin{acknowledgments}
We thank Harold U. Baranger for carefully reading the manuscript and providing multiple comments. Development of the theoretical expressions and data analysis by L.Z., and G.F., and transport measurements by L.Z. and T.L.  were supported by the Division of Materials Sciences and Engineering, Office of Basic Energy Sciences, U.S. Department of Energy, under Award No. DE-SC0002765. Sample fabrication and characterization by L.Z., and E.A. were supported by NSF award DMR-2004870. 
The deposition of MoRe performed by F.A. was supported by a URC grant at Appalachian State University.
K.W. and T.T. acknowledge support from the Elemental Strategy Initiative conducted by the MEXT, Japan, (grant no. JPMXP0112101001), JSPS KAKENHI (grant no. JP20H00354) and CREST (no. JPMJCR15F3, JST).
The sample fabrication was performed in part at the Duke University Shared Materials Instrumentation Facility (SMIF), a member of the North Carolina Research Triangle Nanotechnology Network (RTNN), which is supported by the National Science Foundation (Grant ECCS-1542015) as part of the National Nanotechnology Coordinated Infrastructure (NNCI).

\end{acknowledgments}

\section*{Appendix}

As an example, here we consider the extension of eq.~\ref{eff_distr} to the situation where a normal contact 1 is followed by two superconducting contacts, 2 and 3. The electron distribution downstream of contact 2 is given by Eq.~\ref{eff_distr}:
$f(E)=P_{2,e} f_0(E-eV_1)+P_{2,h} f_0(E+eV_1 - 2eV_2) + (1-P_{2,e}-P_{2,h})f_0(E-eV_2)$, with an effective $\mu_2=(1-P_{2,e}+P_{2,h}) eV_2+(P_{2,e}-P_{2,h})eV_1=(1-P_2) eV_2+P_2eV_1$.

Upon reflections and partial absorption/emission by contact 3, the distribution becomes: 
\begin{widetext}
\begin{equation}
\begin{split}
f(E)&=P_{2,e}P_{3,e} f_0(E-eV_1)+P_{2,e}P_{3,h} f_0(E+eV_1 - 2eV_3)+ P_{2,h}P_{3,e} f_0(E+eV_1 - 2eV_2) +\\& P_{2,h}P_{3,h} f_0(E -eV_1+ 2eV_2-2eV_3) +P_{3,e}(1-P_{2,e}-P_{2,h})f_0(E-eV_2) +\\& P_{3,h}(1-P_{2,e}-P_{2,h})f_0(E+eV_2-2eV_3)+ (1-P_{3,e}-P_{3,h})f_0(E-eV_3)
\end{split}
\label{3rd_cont}
\end{equation}
\end{widetext}
Here, the first four terms describe the normal and Andreev reflections of the electron distribution coming from contact 1; the next two terms describe  electrons coming from contact 2 being reflected by contact 3; and the last term describes the electrons originating in contact 3. Integrating over the energy and simplifying the expression, the effective chemical potential of this distribution acquires a very natural form: $\mu_3=(1-P_3) eV_3+P_3(1-P_2)eV_2+P_2P_3eV_1$. We can interpret this expression as a result of reflecting the upstream distribution by successive contacts, e.g. the $ P_2P_3eV_1$ term corresponds to the electrons coming from contact 1 being reflected by contacts 2 and 3.

Furthermore, we can rewrite this expression as $\mu_3=(1-P_3) eV_3+P_3\mu_2$. Qualitatively, this means that for the purpose of tracking the current, the electrons arriving at contact 3 with an effective chemical potential $\mu_{2}$ are reflected to the electrons leaving contact 3 with a probability $P_3$.

\bibliography{LB}

\end{document}